\documentclass[sigplan,natbib=false,nonacm]{acmart}

\renewcommand\footnotetextcopyrightpermission[1]{}
\usepackage{ifthen}

\newboolean{anon}
\setboolean{anon}{false}

\microtypecontext{spacing=nonfrench}

\usepackage{lmodern}

\usepackage{tcolorbox}
\newtcolorbox{evaluationq}{
    center,
    colframe=gray!90,
    colback=gray!30,
}

\newcommand{\evalq}[2]{
    \begin{evaluationq}
        \textbf{Q:} \emph{#1}

        \textbf{A:} #2
    \end{evaluationq}
}

\usepackage{csquotes}

\usepackage{amsmath}

\usepackage[USenglish]{babel} %

\usepackage{pifont}

\usepackage{listings}
\usepackage[
    backend=biber,
    sorting=ynt,
    maxnames=1000, %
    backref=true,
]{biblatex}
\appto{\biburlsetup}{}%

\newcommand{\cmark}{\ding{51}}%

\NewDocumentCommand{\LSKV}{}{\ifthenelse{\boolean{anon}}{CKVS}{LSKV}}
\NewDocumentCommand{\LSKVfullstore}{}{\ifthenelse{\boolean{anon}}{Confidential Key-Value Store}{Ledger-backed Secure Key-Value datastore}}

\newcommand{\repo}{\ifthenelse{\boolean{anon}}{\emph{redacted}}{\url{https://github.com/microsoft/LSKV}}}

\usepackage{float}

\usepackage{xcolor}
\usepackage{colortbl}

\usepackage[justification=centering]{caption}
\usepackage{subcaption}

\usepackage{todonotes}

\usepackage{ctable}

\newcommand{\Range}{\texttt{Range}}
\newcommand{\Put}{\texttt{Put}}
\newcommand{\DeleteRange}{\texttt{DeleteRange}}
\newcommand{\Txn}{\texttt{Txn}}
\newcommand{\LeaseGrant}{\texttt{LeaseGrant}}
\newcommand{\LeaseRevoke}{\texttt{LeaseRevoke}}
\newcommand{\LeaseKeepAlive}{\texttt{LeaseKeepAlive}}
\newcommand{\Watch}{\texttt{Watch}}

\ifthenelse{\boolean{anon}}{
    \title{CKVS:\@ A Confidential Distributed Datastore to Protect Critical Data in the Cloud}
}{
    \title{LSKV:\@ A Confidential Distributed Datastore to Protect Critical Data in the Cloud}
}

\author{Andrew Jeffery}
\orcid{0000-0003-0440-0493}%
\authornote{Work done while at Microsoft Research}
\affiliation{
    \institution{University of Cambridge}
    \country{United Kingdom}
}
\email{andrew.jeffery@cst.cam.ac.uk}

\author{Julien Maffre}
\affiliation{
    \institution{Microsoft Research}
    \country{United Kingdom}
}

\author{Heidi Howard}
\orcid{0000-0001-5256-7664}%
\affiliation{
    \institution{Microsoft Research}
    \country{United Kingdom}
}
\email{heidi.howard@microsoft.com}

\author{Richard Mortier}
\orcid{0000-0001-5205-5992}%
\affiliation{
    \institution{University of Cambridge}
    \country{United Kingdom}
}
\email{richard.mortier@cst.cam.ac.uk}

\begin{abstract}

Software services are increasingly migrating to the cloud, requiring trust in actors with direct access to the hardware, software and data comprising the service. %
A distributed datastore storing critical data sits at the core of many services; a prime example being \emph{etcd} in Kubernetes.
Trusted execution environments can secure this data from cloud providers during execution, but it is complex to build trustworthy data storage systems using such mechanisms. %
We present the design and evaluation of the \LSKVfullstore{} (\LSKV{}), a distributed datastore that provides an etcd-like API but can use trusted execution mechanisms to keep cloud providers outside the trust boundary.
\LSKV{} provides a path to transition traditional systems towards confidential execution, provides competitive performance compared to etcd, and helps clients to gain trust in intermediary services. %
\LSKV{} forms a foundational core, lowering the barriers to building more trustworthy systems. %

\ifthenelse{\boolean{anon}}{}{Available at \repo{}.}

\end{abstract}

\begin{document}
\maketitle

\section{Introduction}

Distributed datastores are relied upon to store critical data at the core of business-critical applications, emphasizing the need for secure operation.
A popular example is etcd~\cite{etcd} as used in the core of the Kubernetes orchestration platform~\cite{kubernetes}.
All cluster state, including configuration and secrets, is stored in a single etcd cluster~\cite{k8ssecrets, operatingetcdfork8s}.
Attackers with access to the state in the etcd cluster can manipulate resources to cause arbitrary behaviour in Kubernetes.
Since etcd forms the core of the flow of requests within Kubernetes~\cite{rearchk8s} it must provide high performance, correctness, and reliability.

As datastores are increasingly run in the cloud, the data they contain is left unsecured, despite best-practices and encryption in-transit and in-storage.
Datastores and services are increasingly deployed to cloud datacenters due to flexibility and ease of operations.
However, the cloud providers operating the datacenters are not without security incidents themselves~\cite{azurestackrce,azurestackspoofing,azureservicefabricelevation,awscloudformationissue}.
Gaining privileged access to machines provides malicious actors the opportunity to bypass customers' security mechanisms and read data right out of the hardware.

Confidential services can be operated in the public cloud using Trusted Execution Environments (TEEs)~\cite{tees}.
TEEs such as Intel SGX~\cite{sgx}, Intel TDX~\cite{tdx}, AMD SEV-SNP~\cite{sev-snp}, Arm TrustZone~\cite{trustzone} and Arm Realms~\cite{realms} provide the hardware facilities necessary to support confidential computing.
Confidential computing protects data and code in-memory using attested TEEs, preventing unauthorized access or modification during execution, even if the attacker has privileged access to the machine~\cite{confidentialconsortium, towardconfidentialcomputing}.
Newer Intel processors feature more memory for SGX enclaves~\cite{intelbigxeon}, removing the historical limitations of running larger applications in TEEs.
Additionally, Intel TDX and AMD SEV-SNP have support running entire VMs confidentially, confidential VMs~\cite{cvms, cvmgcp, cvmaws, cvmazure}, providing a new avenue for running larger systems in confidential environments.

Despite the new support for running VMs in TEEs, performing a lift-and-shift of existing applications is not the complete story to make them fit into this new threat model.
Continuing to trust the host can lead to the applications' guarantees to being broken.
However, new systems designed for TEEs are not trivial to build.
Work tackling aspects of building on TEEs has been presented covering untrusted host time~\cite{tlease} and storage~\cite{teekv} but they are still challenging to combine together into systems.
Additionally, the applications themselves are complex, requiring consensus~\cite{raft, raftrefloated} and other mechanisms to be correct for proper functioning.

Despite performing a lift-and-shift operation, existing systems still lack adaptations to the new threat model they operate in.
For instance, they may not give end clients a means of validating the operations performed by an intermediate server, such as the Kubernetes API server, leaving them requiring blind trust.

This work presents the \LSKVfullstore{} (\LSKV{}).
\LSKV{} provides confidential operation with an etcd-like KV API including range queries, transactions, leases and watches.
It provides a secure foundation, lowering the barriers to building trustworthy systems.
The contributions of this work are:

\begin{enumerate}
    \itemsep0em

    \item Motivating why existing datastores are not suitable for simple lift-and-shift operation, \textsection{\ref{sec:motivation}}.

    \item A route to transition to confidentiality with \LSKV{}, avoiding the downsides of lift-and-shift, \textsection{\ref{sec:overview}}.

    \item New primitives for waiting for optimistic requests to be processed and enabling clients to gain trust in intermediary services, \textsection{\ref{sec:implementation}}.

    \item \LSKV{}'s competitive and, for some workloads, improved performance over etcd, \textsection{\ref{sec:evaluation}}.

\end{enumerate}

\newcolumntype{C}{>{\centering\arraybackslash}m{1.8cm}}
\ctable[
    caption = Overview of etcd deployment strategies. \LSKV{} provides all the desired features with a smaller Trusted Computing Base (TCB).\@  HW:\@ Hardware; O:\@ Operator; OS:\@ Operating System.,
    label = tab:etcd-progression,
    star,
]
{lCcCCl}
{
    \tnote[1]{Only values are encrypted, not keys or other data.}
    \tnote[2]{Only keys and values are encrypted, not other data.}
    \tnote[3]{Range queries would be possible if using order-preserving encryption.}
}
{
    \toprule
    System                        & Encrypted memory   & Range queries & Proof of writes & Rollback protection & TCB                             \\
    \midrule
    etcd                          & & \cmark{} & & & HW + O + OS \\
    etcd + client V encryption & \cmark{}\tmark[1] & \cmark{} & & & HW + O + OS \\
    etcd + client KV encryption & \cmark{}\tmark[2] & \tmark[3] & & & HW + O + OS \\
    etcd + confidential VM        & \cmark{} & \cmark{} & & & HW + O + OS \\
    \rowcolor{lime!50}
    \LSKV{} on Virtual                       & & \cmark{} & \cmark{}       & & HW + O + OS \\
    \rowcolor{lime!50}
    \LSKV{} on SGX                       & \cmark{} & \cmark{} & \cmark{}       & \cmark{}            & HW \\
    \bottomrule
}

\section{Motivation}\label{sec:motivation}

\paragraph{etcd background}

etcd is \enquote{A distributed, reliable key-value store for the most critical data of a distributed system}~\cite{etcd}.
It provides a comprehensive API, primarily over gRPC~\cite{grpc}, starting from a basic single key-space key-value model with transactions, leases and watches to higher-level primitives such as distributed locks and elections.
It uses the Raft~\cite{raft} consensus protocol to provide strong consistency and durability of its data.
etcd clusters maintain a global revision counter that linearizes operations and can be used for historical queries.

etcd is widely used as a core building block to store critical data in production systems such as Kubernetes, Rook~\cite{rook}, CoreDNS~\cite{coredns}, and M3~\cite{m3}.
This makes the core API a stable, well adopted target to rely upon.
The performance of etcd is also an important aspect of its adoption along with its reliability, providing low maintenance overhead~\cite{etcdwhy}.
Moreover, the protobuf~\cite{protobuf} format used in its gRPC API makes it extensible whilst keeping backwards compatibility.

etcd is run in cloud and on-premise environments, Table~\ref{tab:etcd-progression} outlines some deployment configurations and their properties.
Ordinarily, etcd provides encryption of data in-transit, via TLS connections, and defers encryption of data at-rest to the underlying filesystem~\cite{etcdstorageencryption}.
As the memory is unencrypted, this leaves etcd deployments in the cloud vulnerable, given that the encryption keys reside in-memory.
Clients that do not trust etcd with the confidentiality of their data can encrypt values themselves before sending them to etcd, known as client-side encryption~\cite{etcdhardening}.
Keys can be encrypted with order-preserving encryption~\cite{ope} to retain the ability to perform \Range{} queries.
However, this merely moves security and key-management concerns from the cluster operators to the clients, adding more complexity.

\paragraph{Lift-and-shift}
In order to provide confidentiality of data during execution etcd may be run in confidential VMs: the lift-and-shift approach.
Whilst this provides a simple solution to securing keys and values during execution, the trust model of etcd itself remains.
etcd's trust model relies heavily on the host OS.\@
This trust model leaves it vulnerable to host-controlled attacks such as rollbacks, these attacks have been explored in the context of Engraft~\cite{engraft}, focusing on the Raft protocol which etcd builds around.
For instance, flushing writes to disk should not be on the critical path as the host can respond maliciously, invalidating durability guarantees.
Shims could be used to add some level of rollback protection but they all have downsides in the form of performance impacts, complexity or overheads~\cite{rote, narrator, nimble}.
Thus, a lift-and-shift of etcd can break durability guarantees, making etcd not suitable to be run in confidential environments.

\paragraph{Untrusted API servers}
Since etcd clusters store sensitive state, attackers with the ability to manipulate the values can perform arbitrary operations in a Kubernetes cluster.
This could lead to running malicious workloads to exfiltrate data and disrupt services.
Aside from attacking etcd directly, since clients interact with etcd through the API servers, this exposes another attack vector.
An attacker could control an API server and mutate requests from the client to perform arbitrary operations under the guise of the client.
This would be difficult for the clients to notice, particularly when the attacker ensures a consistent view of the system is presented to the clients.

\section{Overview}\label{sec:overview}

\LSKV{} is a distributed key-value data store for securing confidential data in the cloud, built on the Confidential Consortium Framework (CCF)~\cite{ccf}.\@
It offers API compatibility with etcd with adaptations to fit \LSKV{}'s threat model.
It provides solutions for untrusted intermediaries that terminate TLS connections, as well as an incremental adoption model, to aid users transitioning to confidential datastores in the cloud.

\subsection{CCF}

CCF is a framework for building distributed, highly-available, confidential applications.
It provides application developers with key-value maps for storing state in a ledger and dispatches requests to the application logic based on a REST API model.
The integrity of the ledger is guaranteed by a Merkle Tree~\cite{merkletree}, periodically signed by the current leader node.
The ledger is shared across nodes, replicated using a protocol based on a variant of Raft, requiring signatures of the Merkle Tree root to be replicated before values are considered committed.
Application nodes can run on either a virtual TEE or Intel SGX.\@
The virtual TEE is not confidential and can be run in on-premise production environments where operators are trusted.
SGX is the confidential production TEE, supporting confidential operation and remote attestation, suitable for running in the cloud.

\LSKV{} is an application built on CCF, leveraging its features, but several contributions from \LSKV{} have been upstreamed as part of this work.

\subsection{Data model and API}

\newcolumntype{o}{>{\columncolor{lime!50}}c}
\ctable[
    caption = API outline.,
    label = tab:api,
]{lco}{
    \tnote[1]{Requires a patched CCF}
}{
    \toprule
    RPC            & etcd     & \LSKV{}     \\
    \midrule
    \Range{}          & \cmark{} & \cmark{} \\
    \Put{}            & \cmark{} & \cmark{} \\
    \DeleteRange{}    & \cmark{} & \cmark{} \\
    \Txn{}            & \cmark{} & \cmark{} \\
    \midrule
    \LeaseGrant{}     & \cmark{} & \cmark{} \\
    \LeaseRevoke{}    & \cmark{} & \cmark{} \\
    \LeaseKeepAlive{} & \cmark{} & \cmark{}\tmark[1] \\
    \midrule
    \Watch{}          & \cmark{} & \cmark{}\tmark[1] \\
    \midrule
    \texttt{Receipts}       & & \cmark{} \\
    \bottomrule
}

The \LSKV{} API mimics that of etcd, aiming for wire-compatibility, but includes extensions: the addition of fields to response headers and the addition of a write receipt endpoint.
Table~\ref{tab:api} outlines the API.\@
\LSKV{} accepts requests over either HTTP with JSON payloads or gRPC with protobuf payloads.
This enables flexibility in how applications interact with \LSKV{} from the outset without requiring extra dependencies.

\LSKV{} maintains a single key-space.
Updates to the key-space are versioned with a \emph{revision} counter, incremented for each update.
The revision can be used to query the store at a historical point in time (historical reads).
Response values feature the revision that they were created at (\verb|create_revision|), last modified at (\verb|mod_revision|), and the number of updates to the value since creation (\verb|version|).

Values can have associated leases for tracking client liveness and distributed coordination such as leader election.
The lease is created by a client and is assigned a time-to-live, which the client can refresh.
A lease can be associated with multiple keys and when the lease expires or is revoked the keys will be deleted.
A lease expires if the time-to-live passes without being refreshed, and can be manually revoked by clients.
As there is no way to reliably schedule work in the TEE we perform the deletion of keys with expired leases during a compaction call.
In the meantime, after expiration but before a compaction, leases are soft-deleted --- they will seem to be expired from the client's perspective but still retain storage.

Clients are also able to watch values in \LSKV{}, staying up-to-date without polling.
They can start watching from the latest revision and be streamed updates to specified keys as they occur.
Alternatively, a client can start watching from a historical revision, for instance if the client had to restart but has some stored data and needs to catch-up from a known point.
\LSKV{} only sends updates to clients for values that have been committed in the cluster.
Due to current limitations in CCF for bidirectional HTTP2 streams~\cite{ccfhttptracker}, \LSKV{} requires a patched version of CCF for \Watch{} requests to work.

All responses from the \LSKV{} cluster come with a response header, the fields of which are outlined in Table~\ref{tab:responseheaders}.

\ctable[
    caption = Response header fields.,
    label = tab:responseheaders,
]{ll}{
    \tnote[1]{Unique to \LSKV{}}
}{
    \toprule
    Name            & Description     \\
    \midrule
    Cluster ID          & Cluster-wide identifier \\
    Member ID          & Per-node identifier \\
    Raft term & Latest Raft term  \\
    Revision & Latest revision \\
    Committed Raft term\tmark[1] & Raft term of last commit \\
    Committed revision\tmark[1] & Revision of last commit \\
    \bottomrule
}

\subsection{Threat model}\label{sec:threatmodel}

\begin{figure}
    \centering
    \includegraphics[width=0.8\linewidth]{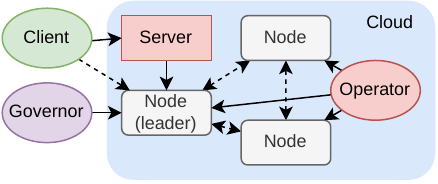}
    \caption{High-level view of a typical 3-node cluster.}\label{fig:highlevel}
\end{figure}

\LSKV{} has three categories of actors, inherited from CCF:\@ operators that manage the running of the application instances, governors that are responsible for management of the running service based off of a JavaScript constitution containing available actions, and clients that call application endpoints, outlined in Figure~\ref{fig:highlevel}.

\textbf{Operators} are \emph{untrusted}, typically being a cloud operator when deploying \LSKV{} to the cloud, and are assumed to have complete control over the host running the application instance.
They can perform denial of service attacks against the \LSKV{} service by turning machines off, or interfering with network traffic.
\LSKV{} does not mitigate these attacks and so cannot maintain liveness in these cases.
Additionally, \LSKV{} does not obfuscate access patterns, mitigate timing attacks, or mitigate other side-channel attacks.
\LSKV{} mitigates operators interfering with reads and writes to storage by not relying on the data to be persisted as part of the guarantees it provides, notably protecting against storage rollback attacks.
Persisted data is encrypted with keys stored in the TEE and so is not readable by the operator, only the governors can get the key to decrypt.
\LSKV{} uses host time for leases and does not mitigate against the time moving forwards abnormally, however time is limited to be monotonically increasing during a node's lifetime.
This is a known limitation of the system and would require support in CCF to work around.
When deployed to a system with a secure TEE \LSKV{} makes standard assumptions about running in a TEE, particularly that code is integrity protected and memory is encrypted and integrity protected.
For SGX there are a number of vulnerabilities~\cite{sgxfail}, the compile-time mitigations are applied to \LSKV{} where available.
Attested TLS is used for node-to-node communication to ensure peers are running in TEES and using TLS for client-to-node communication.

\textbf{Governors} are \emph{trusted in aggregate}: they propose actions from the constitution and these are voted on by other governors.
A proposal must pass a vote threshold before being applied, configurable in the constitution.
The actions available to governors surround node cluster membership, governor membership, service management (opening the service, rotating certificates), and recovery of the service.
\LSKV{} provides a simplified constitution enabling single-governor actions for simplicity but this is configurable.
All governance interactions are signed and available publicly in the ledger.

\textbf{Clients} are \emph{untrusted} apart from using the application endpoints and other read-only endpoints that do not expose sensitive information.
We assume an open security model for clients for simplicity: those that can provide a valid client certificate for the service can use all the functionality, including reading and writing any data in the store.

\subsection{Consistency model}

\LSKV{} provides linearizable writes and serializable reads.
Writes are acknowledged optimistically, not waiting for commit through consensus, instead leaving the action of waiting for consensus to complete for the client, to ensure linearizability of the write.
Reads can be served at any node and clients can wait for the read value to be committed to ensure serializability, the values may be stale.
Within a TLS session \LSKV{} maintains session consistency and this can be enforced manually, for instance across sessions, using the revision field in the response header.

\subsection{Fault and durability model}\label{sec:durability}

\LSKV{} assumes crash-fault tolerance which is limited to having a majority of cluster nodes being available, otherwise disaster recovery is needed.
Nodes do not operate in a Byzantine manner due to the code integrity protection of the TEE.\@
Since \LSKV{} does not trust the host to persist values to disk, data is not eagerly persisted before responding to clients.
This is a fundamental limitation of the threat model: without trusting the host to persist data we cannot guarantee durability of this form.
This equally applies to lift-and-shift systems which have their durability guarantees broken due to the different threat model applied in this context.
Clients wanting to ensure values are available after restarts of the node they are interacting with should ensure that the transaction for their operation has been committed to a majority of nodes, and thus available in-memory on them.

\subsection{Incremental adoption}

There are two ways \LSKV{} supports incremental adoption: TEE flexibility and write receipts.

\begin{figure}
    \centering
    \begin{subfigure}[b]{\linewidth}
        \centering
        \includegraphics[width=0.8\linewidth]{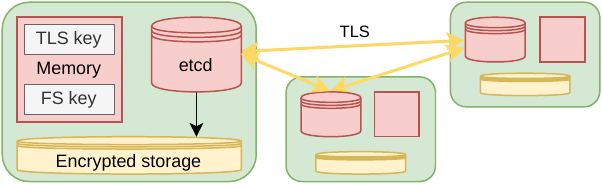}
        \caption{Initial etcd deployment in a private datacenter.}\label{fig:incrementala}
    \end{subfigure}
    \begin{subfigure}[b]{\linewidth}
        \centering
        \ifthenelse{\boolean{anon}}{
            \includegraphics[width=0.8\linewidth]{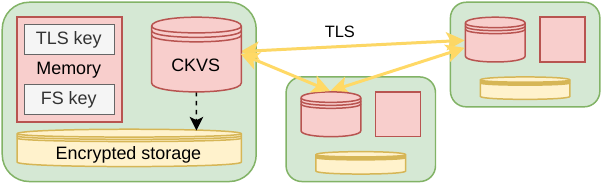}
        }{
            \includegraphics[width=0.8\linewidth]{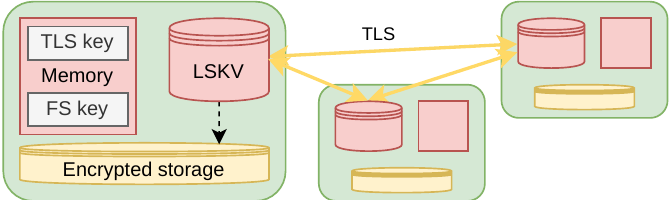}
        }
        \caption{Switching to \LSKV{} in a virtual enclave.}\label{fig:incrementalb}
    \end{subfigure}
    \begin{subfigure}[b]{\linewidth}
        \centering
        \ifthenelse{\boolean{anon}}{
            \includegraphics[width=0.8\linewidth]{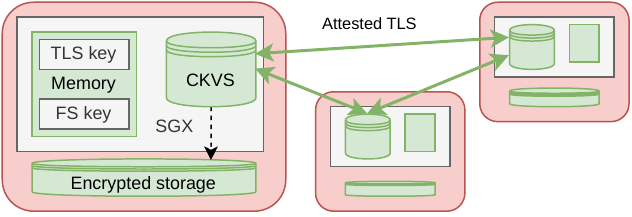}
        }{
            \includegraphics[width=0.8\linewidth]{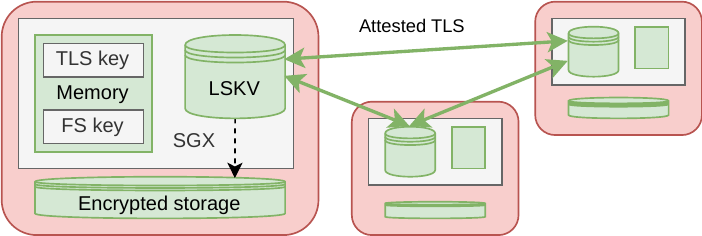}
        }
        \caption{Deploying to a public cloud using \LSKV{} on SGX.}\label{fig:incrementalc}
    \end{subfigure}
    \caption{Architecture and trust during incremental adoption. Green is secure, yellow is using encryption but not necessarily integrity protected, red is insecure. The background represents the security of the environment.}
\end{figure}

\paragraph{TEE flexibility}
Starting from an existing deployment of etcd in a private datacenter, Figure~\ref{fig:incrementala}, we assume that the operator is trusted, TLS is used for network communication and data is being stored on an encrypted disk.
The keys for the TLS communication and filesystem encryption are currently stored in unencrypted memory.
Deploying this configuration to the public cloud, even running etcd in a TEE, would not fit the threat model we want as discussed previously.
Instead, we want to transition the existing service to \LSKV{} incrementally to gain confidence and operational expertise.
Firstly, we make use of TEE flexibility within \LSKV{}, allowing it to run in multiple target environments.
This enables \LSKV{} to be deployed in a virtual TEE, a standard process, in the private datacenter as shown in Figure~\ref{fig:incrementalb}.
This retains the same trust in the operator, and the same conditions for everything else but gives clients a chance to update to any changes required, perhaps waiting for commits.
It additionally gives the operators a chance to test performance, stability and any automated management of their service with it being minimally different from the previous setup.
Later, once operators have confidence in operating the service, they can begin transitioning to a deployment of \LSKV{} in the public cloud using the SGX TEE.\@
This gives the same setup, but now the operator is untrusted, as shown in Figure~\ref{fig:incrementalc}.
Since the operator is untrusted and \LSKV{} is running in a secure TEE it uses attested TLS and the private keys are stored securely in the enclave memory.

\begin{figure}
    \centering
    \ifthenelse{\boolean{anon}}{
        \includegraphics[width=\linewidth]{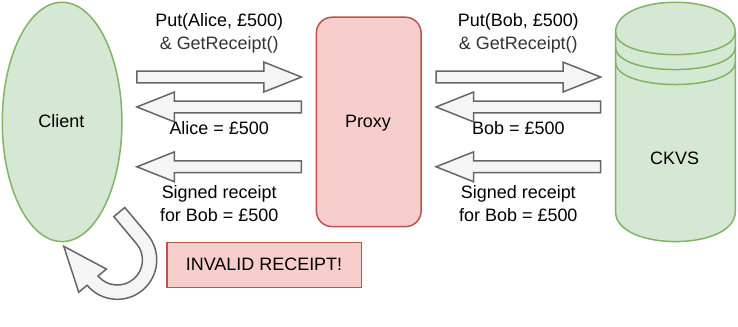}
    }{
        \includegraphics[width=\linewidth]{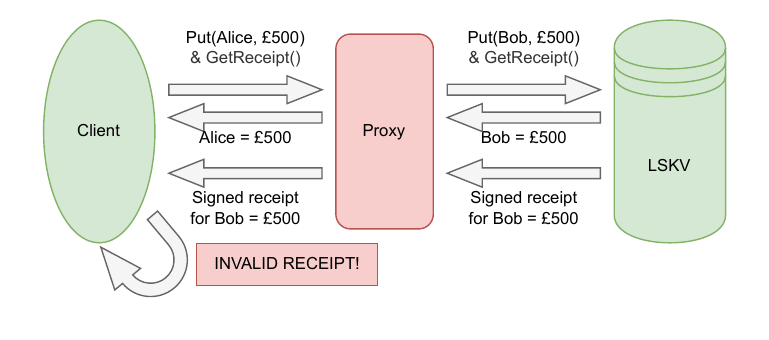}
    }
    \caption{Example of a malicious proxy being detected with write receipts.}\label{fig:proxy}
\end{figure}

\paragraph{Write Receipts}
\LSKV{} provides write receipts for detecting malicious intermediary servers, shown in Figure~\ref{fig:proxy}.
We assume that the server terminates TLS connections and does some intermediate processing on the data.
After performing some request including writes to the intermediate server, clients can request a receipt for the writes.
This receipt provides offline proof that the write was committed to the \LSKV{} cluster and can be used to verify the actions performed by the untrusted server.
The receipt can also be used as proof to other parts of a system that the write request took effect, to ensure that they continue working from a successful state.

\section{Implementation}\label{sec:implementation}

\begin{figure}
    \centering
    \ifthenelse{\boolean{anon}}{
        \includegraphics[width=0.8\linewidth]{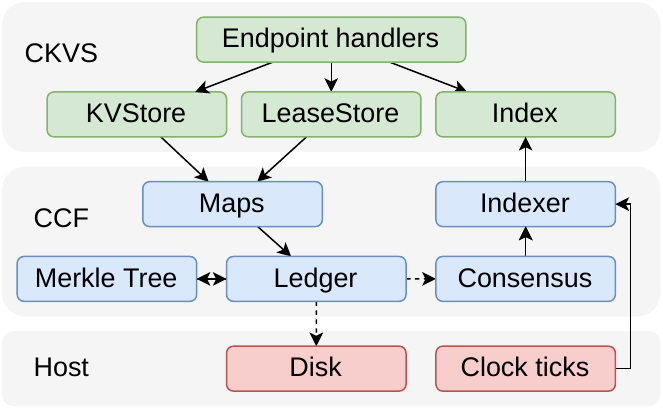}
    }{
        \includegraphics[width=0.8\linewidth]{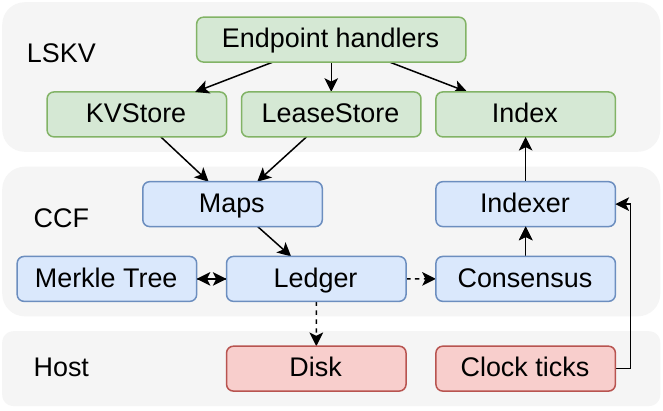}
    }
    \caption{\LSKV{} internals. Dashed arrows imply asynchronous communication.}\label{fig:lowlevel}
\end{figure}

\LSKV{} is implemented as a C++ application on CCF, taking \textasciitilde{}2,100 lines of code, the constitution forms \textasciitilde{}1,200 lines of JavaScript and \textasciitilde{}2,500 lines of code were upstreamed to CCF.\@
Figure~\ref{fig:lowlevel} highlights the separation of functionality offered by CCF and that which \LSKV{} implements.

Requests are routed by CCF and handled by registered endpoint handlers.
These handlers run only on a single thread and perform the primary business logic of updating data in the store using abstractions over CCF maps.
After the handler completes, mutations are stored in the ledger.
When operations get committed they are used to populate the index in \LSKV{}.
This index is then used to serve historical \Range{} requests.

\subsection{Internals}\label{sec:internals}

\paragraph{Response headers}
Each response from \LSKV{} comes with a response header.
The fields contained in a response header are outlined in Table~\ref{tab:responseheaders}.
The cluster ID is a hash of the service's public key for the cluster, only changing for a cluster during disaster recovery.
The member ID is a hash of the node's public key, making it unique to the node that handled the request.
The Raft term along with the revision, a global counter updated with each operation, form the transaction ID for the request.
Transaction IDs identify operations and can be used to check the commit status.
Only requests that mutate the store have an associated transaction ID.\@
Requests that do not mutate the store have a Raft term and revision filled in with the same values as found in the committed Raft term and committed revision, respectively.
The committed Raft term and committed revision form the transaction ID that was last committed at the time of handling the request.
This committed transaction ID is primarily useful to inform the commit status of pending transactions, indicating whether they have been through consensus.

\paragraph{Maps}

\begin{lstlisting}[caption=C++ implementation of a stored value.,label=lst:value]
struct Value {
  std::vector<uint8_t> data;
  int64_t create_revision;
  int64_t mod_revision;
  int64_t version;
  int64_t lease;
}
\end{lstlisting}

Internally, \LSKV{} stores key-value and lease data in CCF maps.
The maps store a byte vector for a key and a JSON serialized \verb|Value| struct (Listing~\ref{lst:value}) as a value.
The \verb|data| field is the bytes of the value that the client sends in a \Put{} request.
The \verb|version| is the number of updates to the value since its creation and the \verb|lease| is the ID of a lease which may be associated with the value.
The \verb|create_revision| is the revision that the value was created at and the \verb|mod_revision| is the revision that the value was last modified at.

When executing a request \LSKV{} operates on an internal CCF transaction which is a snapshot of the key-value store.
However, the transaction's ID is not known until after the execution of the application logic so the revision fields cannot be entered correctly.
Instead, \LSKV{} lazily computes the values of the create and mod revision when loading a value from the map.
On creation of a new value in the map \LSKV{} sets both revisions to 0.
Then, on subsequent operations, the value is first read out of the map and updated to  set the revisions to the correct values.
The map is queried for the ID of the transaction that last modified this key in the map.
The transaction ID's revision is then used to set the create revision, if it was 0, and always set the mod revision of the value.
This means that the revision fields in the values stored in the ledger lag behind by one update.

\paragraph{Consensus and persistence}

Once internal CCF transactions have been executed they are queued for asynchronous replication to other nodes.
Once internal CCF transactions have been replicated to a majority of nodes along with a signature they are deemed committed.
The state of the transaction will then reflect this when queried by clients.
Whilst items are replicated through consensus they are also added to the ledger, encrypted and queued to be persisted to disk asynchronously.
By default \LSKV{} stores all entries in private CCF maps which are stored encrypted in the ledger.

\paragraph{Historical index}

After transactions have been committed with the other nodes they cannot be rolled back in the course of normal operation.
Thus they can be added to a historical index, used for historical \Range{} requests and \Watch{} streams in \LSKV{}.
It is backed by CCF's indexer which periodically applies the latest committed transactions to the historical index, prompted by a tick from the host's clock.
This process is not synchronous with consensus and so the historical index can lag behind the latest committed values.

\paragraph{Public ledger entries}\label{sec:publicledger}

Since \LSKV{} stores all entries in private CCF maps by default, both keys and values are encrypted on-disk.
However, governors may want some keys to be stored unencrypted in the ledger to enable auditability of non-sensitive data.
Governors can alter this by making and accepting governance proposals which are publicly auditable.
Once the proposal is accepted, logic is executed to make new writes to keys with the proposed prefixes publicly readable in the ledger.
On top of these options, clients can still perform their own encryption if the clients have very secret values that they do not trust the governors with, however this should be rare as the governors should be within the trust boundary.

\subsection{Consistency model}\label{sec:consistency}

\LSKV{} is optimistic when processing requests for the latest state, allowing clients to observe values that have not yet been committed, but gives clients the option to be more pessimistic.
It is pessimistic when processing requests for historical values, guaranteeing that readers observe a committed view of the data.
This split consistency mechanism enables the clients to leverage the most useful one to them and their use-case.
In practice this means that:
\begin{enumerate}
    \itemsep0em

    \item After committing mutations, the leader orders the transaction with other executing transactions, assigning it an ID, acknowledges to the client, and then sends the operation through consensus.

    \item The client can check on the status of a transaction ID to wait for it to be committed.

    \item When reading without a revision set, the client may observe values that have not been committed.

    \item Clients can specify a revision to only observe committed values when reading.
\end{enumerate}

\subsubsection{Optimistic (latest data)}

\LSKV{} provides linearizable writes and serializable reads.
These operations are optimistic: they return to the client before waiting for commit.
The writes at the leader node are asynchronously sent to backup nodes through CCF's consensus layer, which performs batching based on configurable count and time intervals.
Meanwhile the client gets a response indicating the revision and Raft term that the write will be present at if it is successfully committed through consensus.
Table~\ref{tab:txstatus} describes the states that a transaction can be in.
With the revision and Raft term, the client can employ different strategies for checking that a write has been committed, outlined below.
Reads can be serviced by any active node in the cluster.

\ctable[
    caption = States of a transaction. Terminal states bolded.,
    label = tab:txstatus,
]{lp{0.7\linewidth}}{
}{
    \toprule
    State & Description \\
    \midrule
    Unknown & Node is unaware of the operation \\
    Pending & Operation is awaiting consensus \\
    \textbf{Committed} & Operation is committed \\
    \textbf{Invalid} & This operation cannot be committed \\
    \bottomrule
}

Since write requests must be served by a leader, requests issued to a non-leader node may be forwarded to the current leader for execution.
Read requests can be served at any active node.

\begin{figure}
    \centering
    \includegraphics[width=\linewidth]{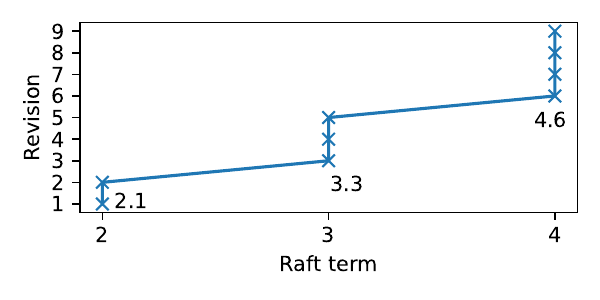}
    \caption{Timeline of Raft term changes and revisions. Annotated vertices show Raft term history entries.}\label{fig:view-history}
\end{figure}

Despite \LSKV{} being optimistic about consistency, some clients may want to wait for values to be committed before continuing.
To support this, \LSKV{} supports methods for checking the status of an operation, given the ID.\@
Clients can use the following strategies to flexibly wait for operations to be committed based on their usage pattern.
Figure~\ref{fig:view-history} shows an example series of transaction IDs and the Raft term history, Table~\ref{tab:commitcheck} summarizes relative performance.

\textbf{Naive}\quad
Poll the transaction status endpoint for each ID until a \emph{terminal} status is obtained for each.
This places extra load on the cluster but makes for simple client logic.

\textbf{Poll last in Raft term}\quad
Locally filter the IDs to the last in each Raft term and apply the naive strategy with these.
If a transaction ID turns out to be invalid then discard it and poll the previous ID for that Raft term.
This strategy is more efficient but requires introspection of the transaction IDs.

\textbf{Poll latest committed transaction}\quad
Poll the latest committed transaction ID in the cluster.
From this ID locally calculate the status of each transaction ID, provided that they are all in the same Raft term.
If a change of Raft term is observed then fall back to one of the previous strategies.

\textbf{Poll latest with Raft term history}\quad
Polling the latest committed transaction can be coupled with the Raft term history, which contains the first transaction ID in each Raft term, to handle Raft term changes efficiently.
This is particularly efficient when the cluster changes Raft terms frequently and it reduces load on the cluster, aiding in faster recoveries.
The Raft term history required for this strategy was upstreamed to CCF as a part of the \LSKV{} work.

\textbf{Using returned committed IDs}\quad
Rather than polling the cluster for statuses and the last committed ID, the ID of the last committed transaction can be used from the response header.
This works best in times of stability, when the Raft term is not changing but can be coupled with periodic refreshes of the Raft term history.
This strategy is most efficient for when making a large number of requests.

\ctable[
    caption = {Comparison of commit checking strategies in terms of number of messages to the service.\ \(n\) is the number of requests waiting for commit, \(t\) is the number of raft term changes that have occurred during the execution.},
    label = tab:commitcheck,
]{lll}{
}{
    \toprule
    Strategy               & Best case     & Worst case   \\
    \midrule
    Naive                  & \(O\)(\(n\))  & \(O\)(\(n\)) \\
    Poll last              & \(O\)(\(1\))  & \(O\)(\(n\)) \\
    Poll committed         & \(O\)(\(1\))  & \(O\)(\(t\)) \\
    Poll with raft history & \(O\)(\(1\))  & \(O\)(\(1\)) \\
    Returned committed     & \(O\)(\(1\))  & \(O\)(\(1\)) \\
    \bottomrule
}

\subsubsection{Pessimistic (historical data)}

Compared to requests operating on the latest state, requests working on the historical state of the store can only observe committed values.
Reads are served from the index which tracks the committed values and does not contain optimistic values.
Separating index updates from consensus rounds keeps them off the hot path, keeping optimistic operations fast at the cost of staleness in historical queries.
Each node maintains their own index so different nodes may have different staleness profiles.
Since every response from \LSKV{} includes the revision and Raft term of latest committed item clients can use this as an indication of the latest state available in the index across nodes they interact with.

\subsection{Auditability}\label{sec:auditability}

When a value is committed in CCF there is a corresponding signature over the internal Merkle Tree state.
This signature is stored in the ledger along with the entries used to form the Merkle Tree.
Since all operations are recorded in the Merkle Tree a valid signature can be used to confirm that an operation was committed.
This signature also identifies the node that created it.
These signatures are stored publicly in the ledger and can be used to validate the ledger by all with access to it.

\paragraph{Write receipts}
Clients may not always be able to connect to \LSKV{} nodes directly, instead interacting with an intermediary such as the Kubernetes API server.
These terminate TLS sessions, potentially aggregating requests to the datastore or presenting their own API.\@
However, clients must now trust the intermediate server to both handle their data safely and faithfully perform their operations.
Preventing the intermediate server from leaking confidential data is out of scope of \LSKV{} but may be mitigated by client-side encryption of values.

To avoid clients having to trust intermediary servers to faithfully perform their requests, \LSKV{} can provide unforgeable write receipts.
These write receipts provide an end client with cryptographic proof to validate that the action it requested the intermediary to perform is what was executed at \LSKV{} and the results of mutations have been committed to the ledger.
To request a write receipt clients submit a revision and Raft term (the transaction ID) of a previous request to a get receipt endpoint.
\LSKV{} then fetches the receipt asynchronously, presenting it to the client once available.
Receipts from \LSKV{} include a digest of the serialized request and response, which the client has possession of and so can verify the receipt themselves.

The structure of a write receipt is outlined in Listing~\ref{lst:receipt}.
The \verb|node_id| is the ID of the node that generated the receipt, \verb|cert| is its public certificate.
Fields under \verb|leaf_components| form a leaf in the Merkle Tree; the \verb|write_set_digest| is a hash of the keys written to during a transaction, \verb|commit_evidence| is a per-transaction string that guarantees the transaction is committed and \verb|claims_digest| is a hash of the custom claims made by \LSKV{}.
\verb|proof| is a list of steps to successively combine with the calculated leaf node to obtain the root of the Merkle Tree.
The \verb|signature| is the signature over the root of the Merkle Tree.
\LSKV{} extends CCF's write receipts by recording the serialized request and response as custom claims when mutating requests are made.
The hash of these claims is used in a receipt to prove that a request was handled, and results of mutations from it are stored in the committed ledger.

Receipt verification is broken into three stages: confirming the claims digest is correct, checking that the receipt is valid, and checking that the signing certificate is trusted.
To calculate the claims digest the client needs to calculate the SHA-256 hash the protobuf serialized request and response, removing the \verb|header| field in the response as it is not filled in during transactions in \LSKV{} and so is not recorded in the claims.
The client should then confirm their calculated value is the same as the receipt-provided \verb|claims_digest|.
To check the receipt's validity the client must rebuild the root of the Merkle Tree.
They should hash the \verb|commit_evidence| field and concatenate the \verb|write_set_digest|, hash of the \verb|commit_evidence|, and the hash of the custom claims to produce the \verb|leaf|.
The \verb|leaf| is then combined successively with the \verb|proof| elements, concatenating the current item to the left or right as given and hashing the result, to calculate the root.
Finally, the client should verify the signature over the calculated root.
To confirm that the node signing the receipt is trusted by the \LSKV{} cluster a client should confirm that the service certificate of the cluster endorses the node certificate given in the receipt.

\begin{minipage}{\linewidth}
    \begin{lstlisting}[caption=Structure of a write receipt in YAML.,label=lst:receipt]
node_id: "..."
cert: "-----BEGIN CERTIFICATE-----..."
leaf_components:
  write_set_digest: "..."
  commit_evidence: "..."
  claims_digest: "..."
proof:
- left: "..."
- right: "..."
signature: "..."
\end{lstlisting}
\end{minipage}

\subsection{Discussion}

\paragraph{Incremental adoption}
For users with current on-premise etcd deployments there is likely to be friction in switching to other offerings due to having to change client-side code, operational infrastructure, as well as simply requiring developers to learn new systems.
The approach \LSKV{} takes to these challenges is to extend current systems, keeping core API compatibility, rather than creating new interfaces.
This means that client-side code needs only minimal changes in order to wait for commit, operational infrastructure needs minimal changes due to the change in threat model, and developers only have to learn minimal new features if they want to use them, which is not a requirement.
This model aims to greatly accelerate the adoption of confidential computing platforms, making them available to the masses.
The approach taken by \LSKV{} to solve this problem is something that can be reflected in further systems design.

\paragraph{Optimistic consistency}
In light of aiding adoption, whilst updating the threat model, \LSKV{} is optimistic about consistency.
Upon response, clients do not have a guarantee of their values being committed, rather they have the option to wait for commit.
It would be feasible to operate an intermediate server between clients and \LSKV{} that waits for values to commit before responding to retain strong consistency by default, reduce the transitioning burden.
This could take a similar role as etcd's gateway~\cite{etcdgateway} that performs other operations, such as watch aggregation, to extend the scalability of the cluster.
Alternatively, it could be built into platforms that have API servers for other functionality.
The flexible waiting primitives that \LSKV{} provides emphasise the opportunities for clients to be in control of their consistency and performance.

\paragraph{Untrusted servers}
Whilst data confidentiality is a primary focus of \LSKV{}, being able to build trust in systems is also a key concern.
Clients making requests to write data into \LSKV{}, whilst trusting the intermediary with the data may want confirmation and a guarantee that data was written into \LSKV{} with a write receipt.
Write receipts can also be passed to other clients as proof that requests were performed and data written back as expected.

\section{Evaluation}\label{sec:evaluation}

To evaluate \LSKV{} we first compare it with etcd, before exploring other factors of \LSKV{}'s performance.
We investigate the following aspects:

\begin{enumerate}
    \itemsep0em

    \item \LSKV{}'s performance compared to etcd~\textsection{\ref{sec:lskvvetcd}}

    \item \LSKV{}'s horizontal scalability~\textsection{\ref{sec:scalability}}

    \item \LSKV{}'s vertical scalability~\textsection{\ref{sec:verticalscaling}}

    \item The impact of optimism~\textsection{\ref{sec:commitsandreceipts}}
\end{enumerate}

\begin{figure*}
    \centering
    \includegraphics[width=0.8\linewidth]{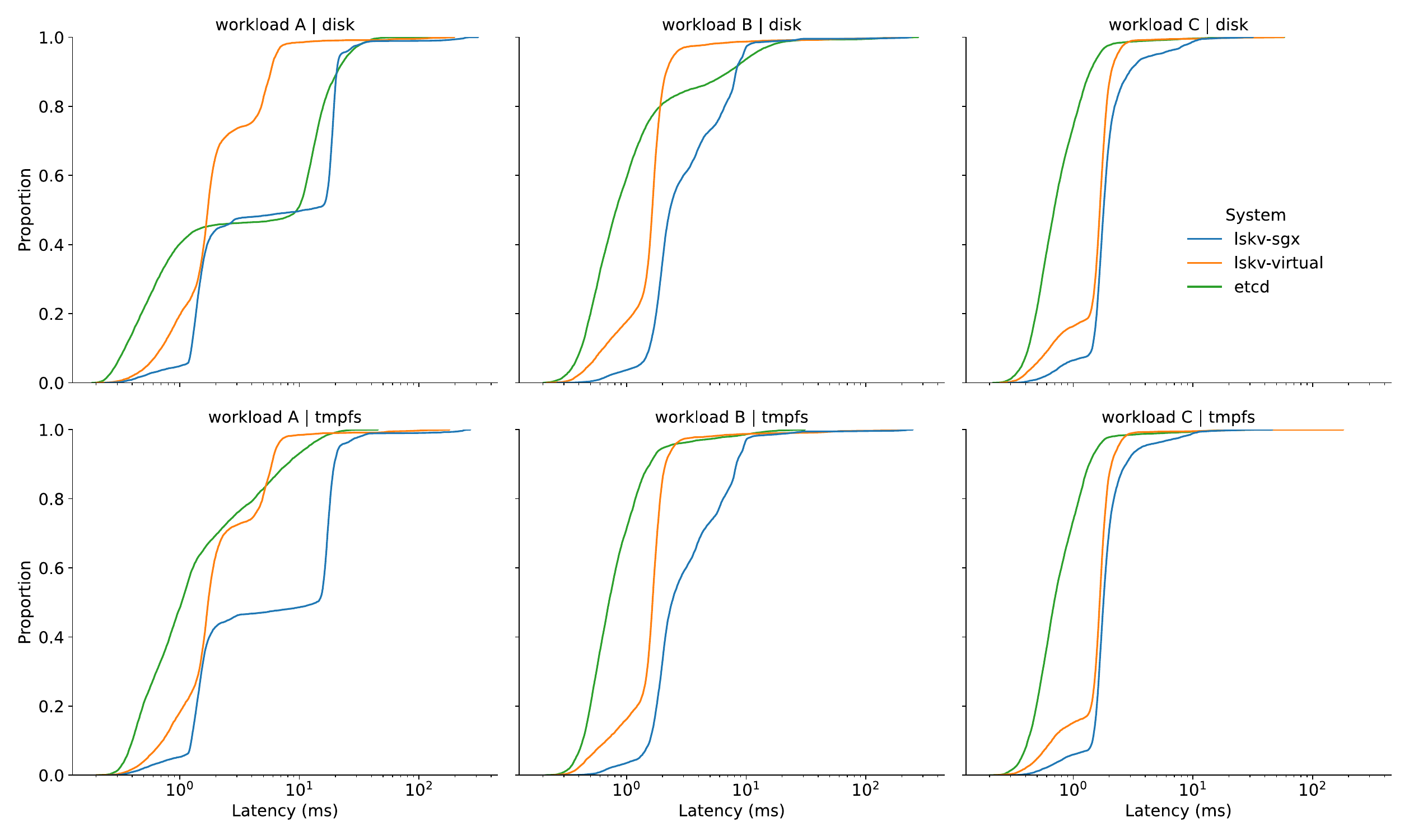}
    \includegraphics[width=0.8\linewidth]{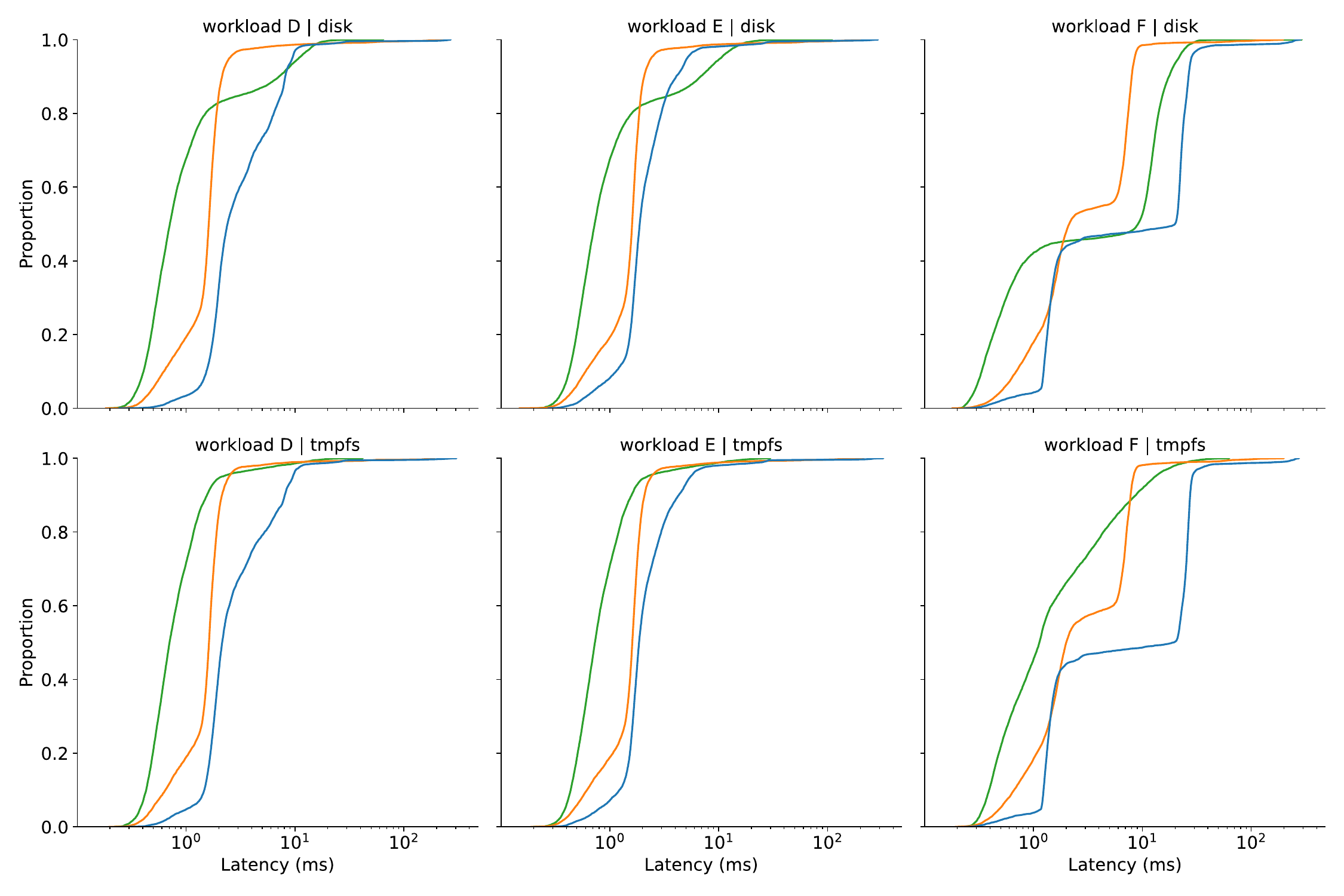}
    \caption{YCSB workloads against etcd and \LSKV{} on disk and tmpfs with 3 nodes, 20,000 requests per second.}\label{fig:ycsbworkloadslatency}
\end{figure*}

\subsection{Setup}

All of the benchmark runs were performed in a cluster of virtual machines in Microsoft's Azure cloud, using the \enquote{East US} location.
All machines in this cluster had the \enquote{Standard\_DC4s\_v3} machine type, which equates to 4 vCPUs, 32GiB memory, with a premium SSD.\@
They were running Ubuntu 20.04 for their OS.\@
The machines have support for spawning Intel SGX enclaves, which are used for \LSKV{} running in SGX mode.
Datastore nodes were run on separate machines, with full access to its resources, in the cluster and the benchmark clients run from a single separate machine in the cluster.
Mutating operations (puts and deletes) target the leader node at the start of the run, read operations target all nodes in a round-robin fashion.
For the SGX enclave build of \LSKV{} the enclave is set with \lstinline|NumHeapPages| equal to 500,000.
Each page is 4KiB so this equates to a maximum of 2GB of heap memory.
The benchmarks were repeated 10 times and the plots presented summarise all the repeats.
\LSKV{} is run with a base configuration of 2 worker threads and a signature interval of 1s.

\begin{figure}
    \centering
    \includegraphics[width=\linewidth]{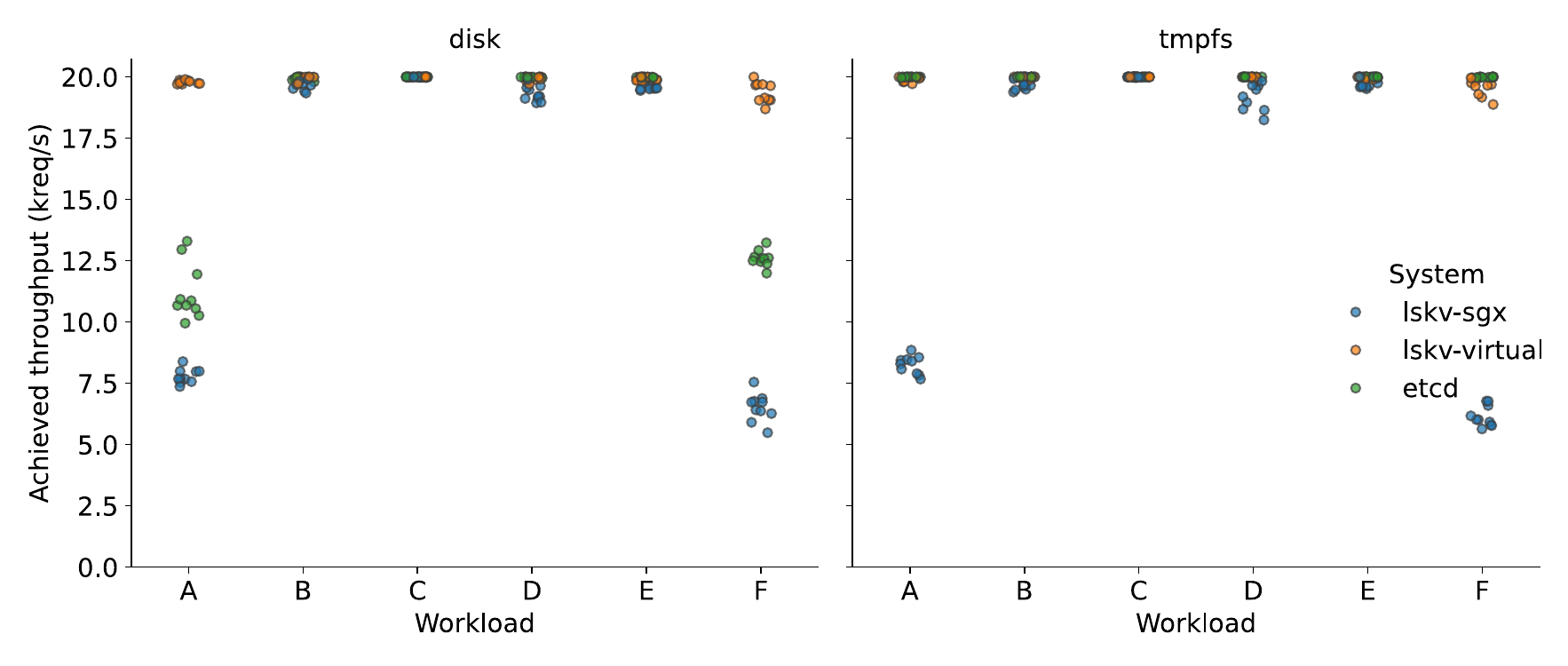}
    \caption{Total throughput of YCSB workloads against etcd and \LSKV{} on disk and tmpfs with 3 nodes, 20,000 requests per second.}\label{fig:ycsbworkloadsthroughput}
\end{figure}

\paragraph{YCSB benchmark}
The Yahoo! Cloud Serving Benchmark (YCSB)~\cite{ycsb} is a standard benchmark for distributed storage systems, presenting workloads based on real-world scenarios.
We use a custom Rust implementation in-place of the original Java version.
For the presented experiments the client uses 100 virtual clients to issue requests for all workloads in a closed-loop fashion.
Tests comparing with etcd target 20,000 requests per second and others target 10,000 requests per second, all running for 10 seconds.
All workloads use a zipfian distribution.
Table~\ref{tab:ycsb-workloads} describes the workloads used.
For \LSKV{}, the writes do not include the time to wait for a commit.
The read-modify-write operation is implemented as a native etcd transaction and all reads are serializable.
We use only workload A after the comparison with etcd as it represents a balanced mix of reads and writes.

\ctable[
    caption = YCSB workload characteristics.,
    label = tab:ycsb-workloads,
]{cl}{
}{
    \toprule
    Workload            & Description     \\
    \midrule
    A & Update heavy (50\% reads, 50\% updates) \\
    B & Read mostly (95\% reads, 5\% updates) \\
    C & Read only (100\% reads) \\
    D & Read latest (95\% reads, 5\% inserts) \\
    E & Short ranges (95\% scans, 5\% inserts) \\
    F & Read-modify-write (50\% reads, 50\% rmw)\\
    \bottomrule
}

\paragraph{Latency measurement}
The latency records the time taken for a node to process a request and respond, measured at the client.
It is calculated from the time recorded at the start of sending the request, and at the end of receiving the response.
This assumes that the connection has already been established and is maintained throughout the run.

\subsection{\LSKV{} vs etcd}\label{sec:lskvvetcd}

\evalq{
    How do the differing internal mechanics of the etcd API exposed by \LSKV{} impact performance?
}{
    \LSKV{} is competitive with etcd.
}

Figure~\ref{fig:ycsbworkloadslatency} shows the latency and Figure~\ref{fig:ycsbworkloadsthroughput} the total throughput results of YCSB workloads applied to \LSKV{}-sgx, \LSKV{}-virtual and etcd version 3.5.4 with 3 nodes.

Presenting the same core API as etcd leads clients of \LSKV{} to expect similar performance characteristics.
However, since \LSKV{} performs more work to offer extra functionality we expect there to be a small overhead.
Since SGX builds of \LSKV{} include extra mitigations we expect this platform to be more severely impacted.
Through all the YCSB workloads \LSKV{} on disk keeps competitive write performance with etcd, reads on etcd are lower latency and when run on tmpfs etcd consistently wins.
All of the datastores are able to attain the applied load rate, apart from \LSKV{}-sgx on workloads A and F which feature higher proportions of writes posing a higher CPU workload.

The writes to \LSKV{} do not wait for commit, as the etcd writes do.
This comes down to a core trade-off in \LSKV{} between commit latency and throughput as producing the commit signatures is costly, explored more in \textsection{\ref{sec:commitsandreceipts}}.
Despite this, the steady-state of these systems places the emphasis on being optimistic, with clients falling back to wait for commit if absolutely necessary.
Leader elections would lead to lower performance as clients may turn more pessimistic, waiting for commits more until \LSKV{} returns to stability.

It is clear to see the distinction between reads and writes for etcd in workloads A and F in the stepped latency when running on a disk.
This is less extreme with a smaller proportion of writes occurring such as in workloads B, D and E and latency significantly improves in workload C due to no writes.
Coupled with the observation that this step is no longer present when running on a tmpfs, this implies that writes in etcd are expensive primarily due to the requirement to flush to disk before returning, in order to guarantee persistence.
\LSKV{}-sgx also sees a step-wise increase in latency for large volumes of writes, however, this continues when running on a tmpfs indicating that the writes are incurring the overhead of cryptography and added mitigations for SGX, as they do not synchronously flush to disk.
For write-heavy workloads, A and F, \LSKV{}-virtual provides a much more consistent experience to clients due to the optimistic consistency model and the lack of need for mitigations and their associated overhead.
Given that \LSKV{} does more work on each request at the leader node, processing the data to the ledger and updating the merkle tree, these results align with our expectations.

Despite the significant impact of the mitigations for SGX newer platforms show that these overheads could be significantly reduced, bringing performance closer to that of the virtual build.
In particular, AMD's SEV-SNP poses an opportunity to run applications in a confidential environment with a lower overhead of 2--8\% compared to virtual, according to a joint analysis by AMD and Azure~\cite{sevsnpenclaveperformance}.

\subsection{Horizontal scalability}\label{sec:scalability}

\evalq{
    How well does \LSKV{} scale horizontally?
}{
    \LSKV{} scales like a typical Raft-based system.
}

\begin{figure}
    \centering
    \includegraphics[width=\linewidth]{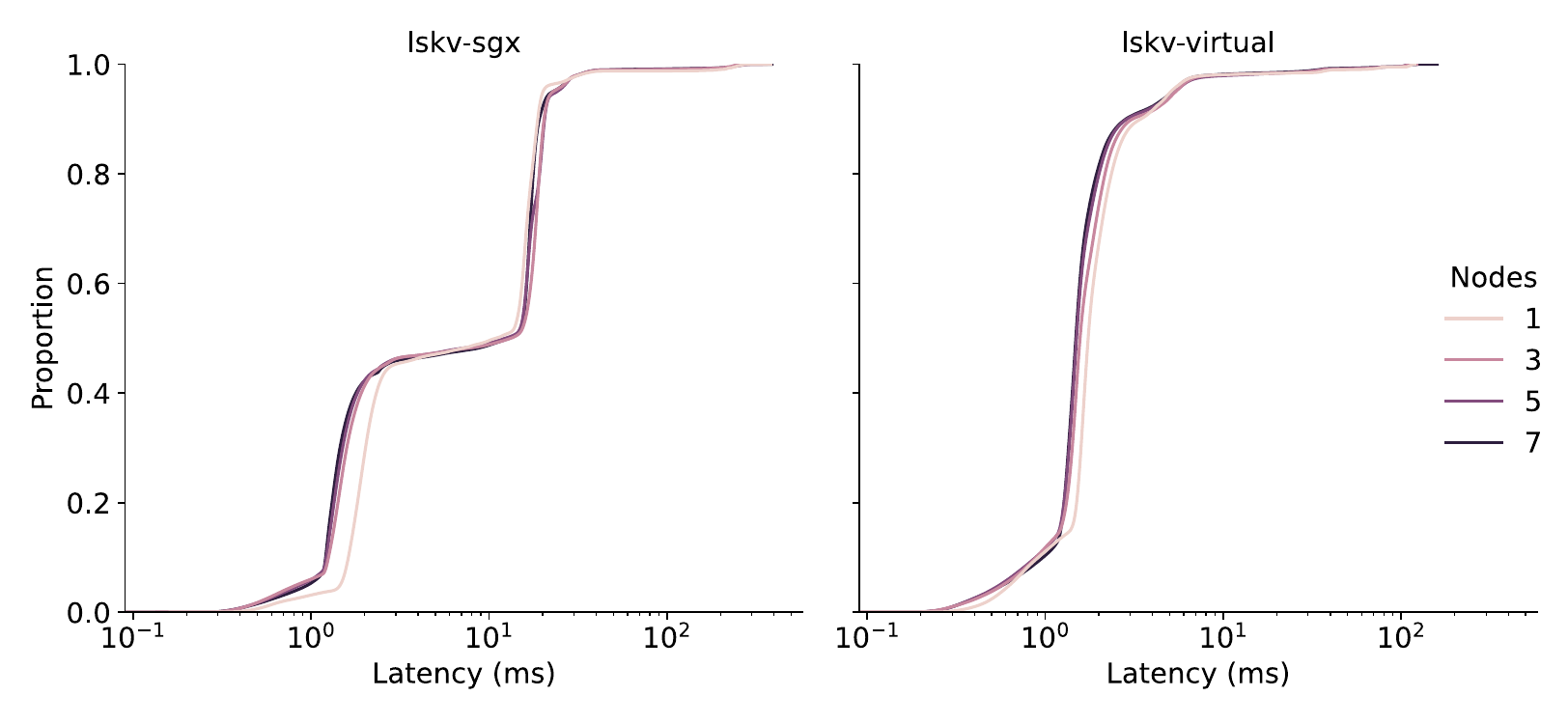}
    \caption{Varying cluster size, 10,000 requests per second.}\label{fig:scalability}
\end{figure}

The scalability of a distributed system is typically important in order to be able to support increased redundancy and attain higher performance.
This experiment, results shown in Figure~\ref{fig:scalability}, exposes the scaling properties of \LSKV{} under the YCSB workload A.
The virtual mode is able to handle the load with a slight increase in latency for one node.
However, for SGX mode the reads are served with expected latency at all scales but the updates experience improved performance on 3 and 5 nodes compared to 1 dropping back down at 7 nodes indicating an overload of the leader with 7 nodes due to the extra replication requirements.

\subsection{Vertical scalability}\label{sec:verticalscaling}

\evalq{
    How well does \LSKV{} scale vertically?
}{
    \LSKV{} benefits from additional parallelism.
}

\begin{figure}
    \centering
    \includegraphics[width=\linewidth]{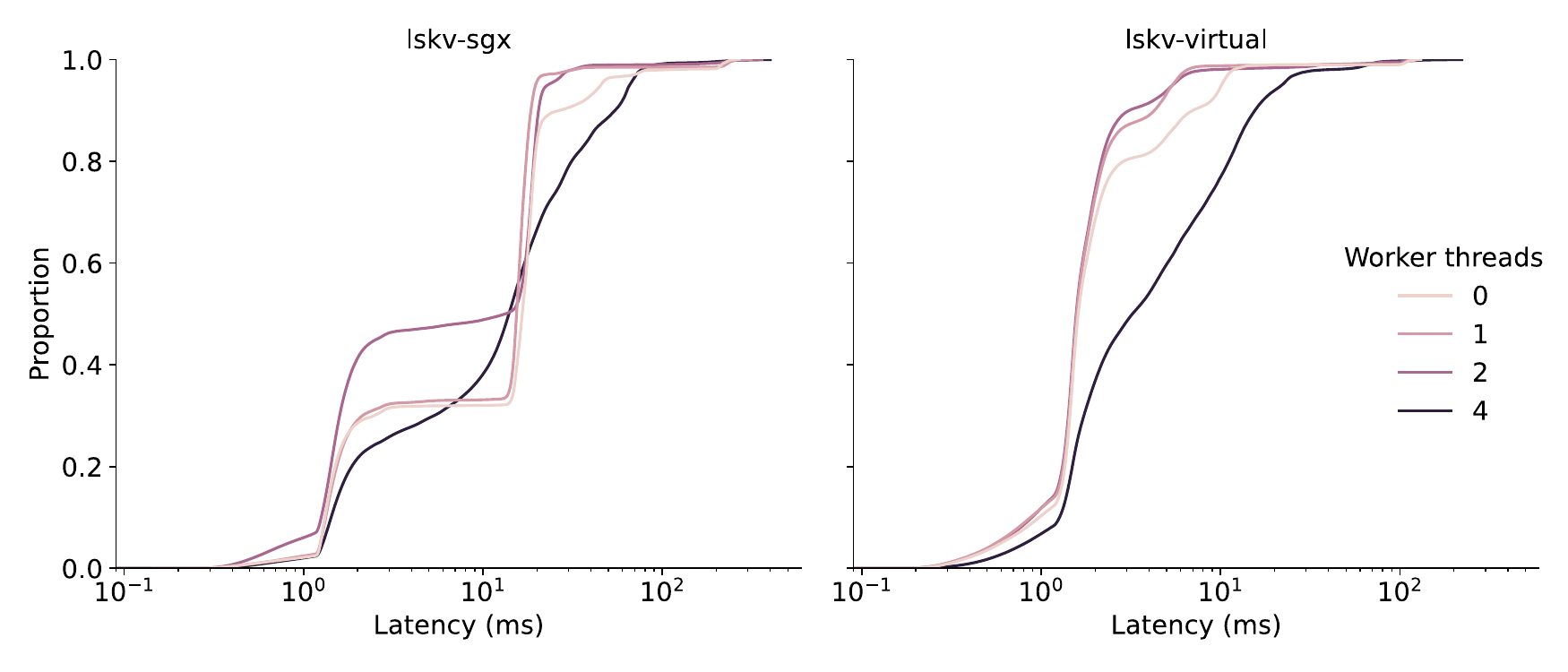}
    \caption{Varying the worker threads, 10,000 requests per second.}\label{fig:workerthreads}
\end{figure}

Figure~\ref{fig:workerthreads} presents results from varying the number of additional worker threads used for a YCSB workload A.
Having 1 additional worker thread from the base of 0 seems to reduce latency, particularly at the tail for updates on both virtual and SGX.\@
An extra worker thread, making 2, also improves latency however matching the number of worker threads to that of the number of cores present on the machine degrades performance.
This is expected due to CCF using a base number of two threads for the main processing of transactions and networking.

\subsection{Commit latency and receipts}\label{sec:commitsandreceipts}

\evalq{
    How does optimistic consistency impact performance and receipt generation?
}{
    The level of optimism directly impacts the commit and receipt delay.
}

\begin{figure}
    \centering
    \includegraphics[width=0.8\linewidth]{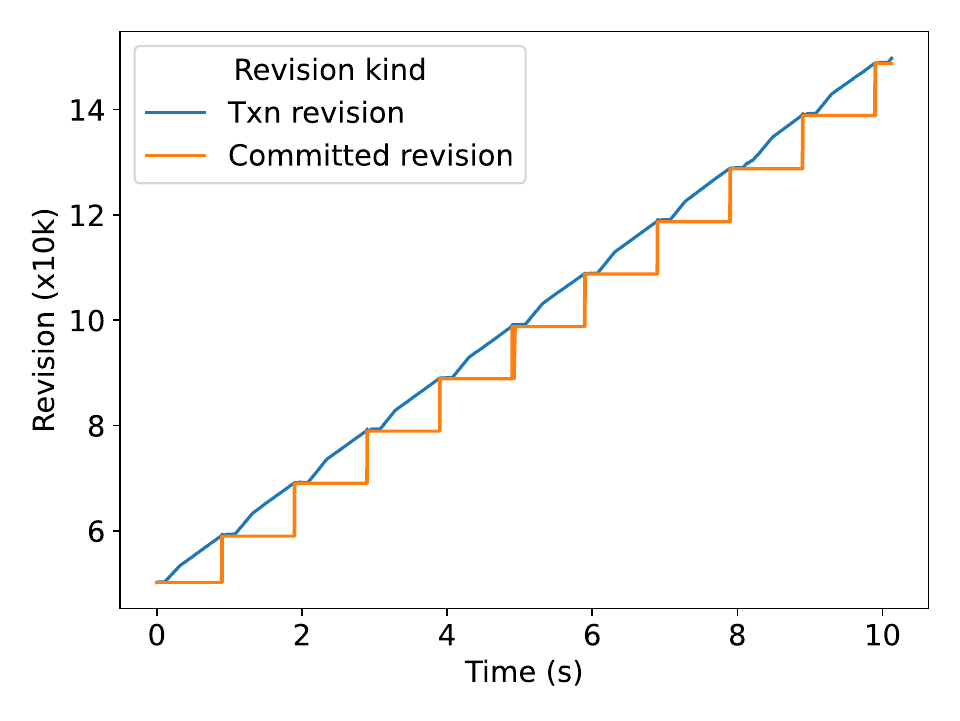}
    \caption{Commit progress during a single YCSB workload A benchmark run.}\label{fig:commitlatency}
\end{figure}

\begin{figure}
    \centering
    \includegraphics[width=\linewidth]{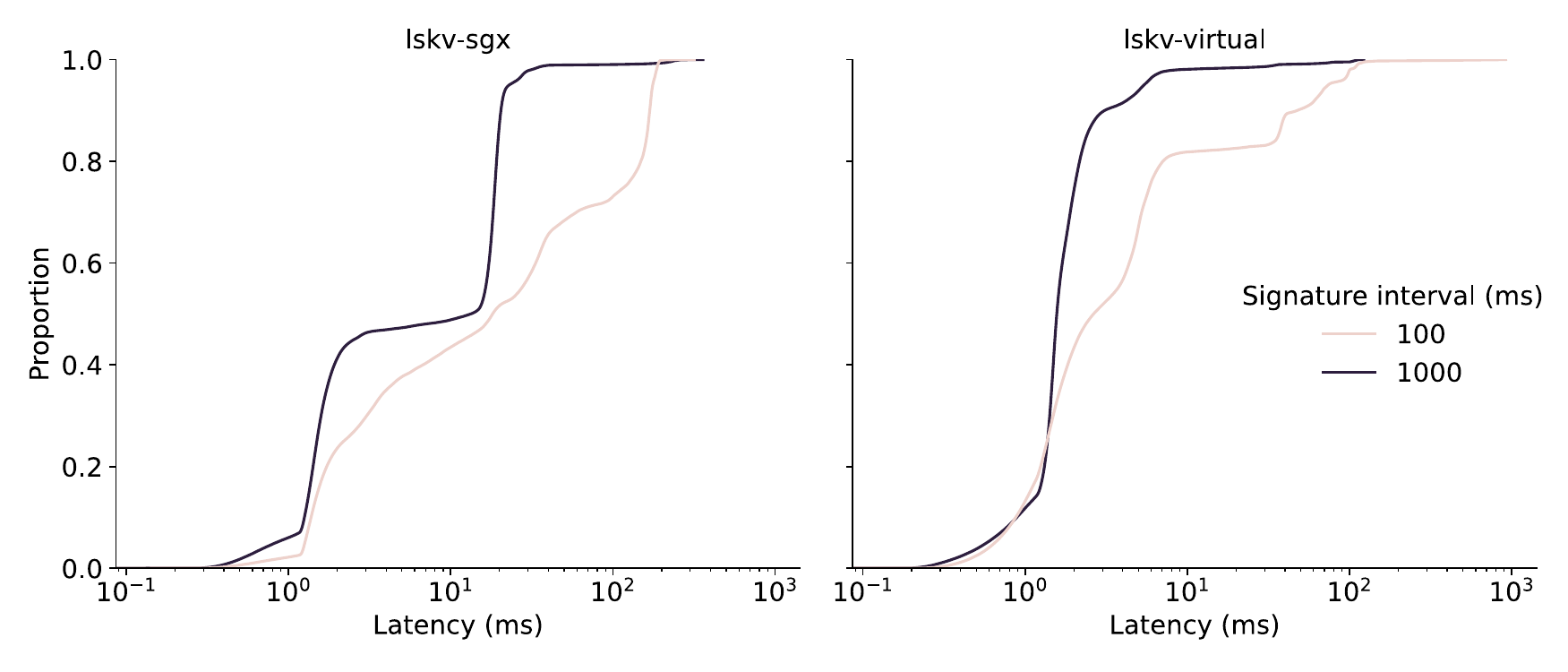}
    \caption{Varying the signature interval, 10,000 requests per second.}\label{fig:siginterval}
\end{figure}

Since \LSKV{} provides optimistic consistency, Figure~\ref{fig:commitlatency} highlights an example of how commits lag behind during a benchmark run.
The commits are seen at 1 second intervals (the vertical jumps of the committed revision), the value set for our evaluation, though this is tunable for deployments.
This means that clients would have to wait at most approximately 1 second before their value gets committed.
The impact of increasing the signature frequency is shown in Figure~\ref{fig:siginterval}, showing an increase in latency of all aspects from the more frequent signatures.
This is because the leader must spend more of its time computing the signature instead of processing transactions.

This also has direct impacts on the latency for obtaining receipts, which require the operations to be committed.
Tuning the signature interval to be more frequent would reduce this latency but add more load to the leader for creating the signatures.
Receipts can be generated by non-leader nodes to aid in handling the extra computation.

Once clients obtain a receipt they need to verify it offline.
To evaluate this we wrote a Python benchmark using the CCF library for validation.
A hard-coded receipt was used, along with service certificate to check the claims were correct, the signature was valid, and that the node certificate was endorsed by the service certificate.
This setup could achieve 541 verifications in sequence per second on a single machine.

\section{Related work}

\paragraph{Embedded datastores}
FastVer~\cite{fastver} extends Faster~\cite{faster}, an embedded concurrent and integrity-protected key-value store, with a \verb|verify| method for data integrity based on a Merkle Tree.
Being embedded, Faster does not offer fault tolerance itself, leaving this to the wrapper program, unlike \LSKV{} that handles fault tolerance and replication natively.
Faster leverages concurrency heavily compared to \LSKV{} which handles core logic on only a single thread.
\LSKV{} internally uses CCF's Merkle Tree which, as demonstrated by FastVer, can alone reach only 100,000 operations per second, working purely in-memory on a single thread on a virtual TEE.\@

ShieldStore~\cite{shieldstore} and Precursor~\cite{precursor} work around the old limitation that SGX enclaves had very limited memory available, however, since this limitation no longer exists regular in-memory data structures can be used.

\paragraph{Confidential distributed building blocks}
T-Lease~\cite{tlease} presents a distributed lease primitive, similar to those provided by \LSKV{}, that works on untrusted time without violating the properties of a lease.
\LSKV{} does not protect the lease properties directly, using the host-provided time instead.
T-Lease would pose a good further extension to \LSKV{}, including generalizing it to cross-platform implementations.

Treaty~\cite{treaty}, Engraft~\cite{engraft} and Enclage~\cite{enclage} all implement components of building distributed confidential applications, covering transactions, consensus, and storage respectively.
Treaty manages distributed transactions over multiple nodes using two-phase commit, whereas \LSKV{} executes transactions on a leader node, replicating the results through a variant of Raft.
Engraft implements Raft over nodes running TEEs, offering a reusable Raft implementation.
This Raft implementation is another variant of Raft compared to CCF's but tolerates the same number of node failures: \(f\) out of \(2f\)+1 nodes.
Enclage implements a performant, encrypted storage engine designed to leverage enclave-native concepts, but does not cover data integrity.
\LSKV{}'s backing ledger stores private data encrypted with a ledger key and persists integrity-protected files to disk.

VeritasDB~\cite{veritasdb} provides a proxy that sits between unmodified clients and existing database servers to guarantee integrity to the client in the presence of exploits or implementation bugs in the database servers.
This is limited to integrity, not full confidentiality of the data, despite the proxy running in an SGX enclave.

\paragraph{Distributed confidential datastores}
Avocado~\cite{avocado} and EdgelessDB~\cite{edgelessdb} are distributed datastores that present different persistence guarantees.
Avocado is in-memory only, similar to \LSKV{}'s optimistic approach, not relying on data to be persisted to disk.
It supports integrity-protection of data and provides strong consistency for client requests.
Avocado does not support transactions, ranges, leases, watch requests or write receipts, and for a comparable YCSB setup achieves similar results to \LSKV{}.
EdgelessDB aims to be compatible with MySQL databases whilst offering confidentiality of data during execution.
It serves requests with multiple cores and eagerly persists data to storage, unlike \LSKV{} but does not support features such as leases, watches and write receipts.
Additionally, it does not support multiple nodes, sacrificing on the availability of the service.

\section{Conclusion}

In this work we have presented \LSKV{}, the \LSKVfullstore{}.
It builds on top of CCF, keeping cloud operators out of the trust boundary when running in confidential TEEs but can be run on-premise outside of a TEE for more performance.
It presents a familiar etcd-like API, easing the transition of existing services to confidential environments.
It provides a consistency model suited to the trust boundary it works within, reducing reliance on the host, unlike common lift-and-shift situations.
It helps clients gain trust in intermediary services with write receipts and achieves competitive performance compared to etcd, in a comparable setting.
Overall, \LSKV{} enables building trustworthy systems to work securely with critical data in the cloud, offering a secure foundation for new confidential systems.

\ifthenelse{\boolean{anon}}{}{
    \section*{Acknowledgments}

    Thanks to the CCF team at Microsoft Research for their help during this project.
}

\printbibliography{}

\end{document}